\documentclass[%
 aip,
 amsmath,amssymb,
 reprint,%
]{revtex4-1}

\usepackage{graphicx}
\usepackage{dcolumn}
\usepackage{bm}
\usepackage[utf8]{inputenc}
\usepackage[T1]{fontenc}
\usepackage{mathptmx}
\usepackage{gensymb}

\usepackage{dcolumn}
\usepackage[normalem]{ulem}

\begin{document}

\preprint{AIP/123-QED}

\title{Experimental investigation of the return flow instability\\in magnetic spherical Couette flow}

\affiliation{
Institute of Fluid Dynamics, Helmholtz-Zentrum Dresden-Rossendorf, Bautzner Landstraße 400, 01328
Dresden, Germany
}

\author{J. Ogbonna}
\altaffiliation{Author to whom correspondence should be addressed: j.ogbonna@hzdr.de}
\author{F. Garcia}
\author{T. Gundrum}
\author{M. Seilmayer}
\author{F. Stefani}

\date{\today}

\begin{abstract}
	We conduct magnetic spherical Couette (MSC) flow experiments in the return flow instability regime with GaInSn as the working fluid, and the ratio of the inner to the outer sphere radii $r_{\rm i}/r_{\rm o}$~=~0.5, the Reynolds number ${\rm Re}$~=~1000, and the Hartmann number ${\rm Ha} \in [27.5,40]$. Rotating waves with different azimuthal wavenumbers $m \in \{2, 3, 4\}$ manifest in certain ranges of ${\rm Ha}$ in the experiments, depending on whether the values of ${\rm Ha}$ were fixed or varied from different initial values. These observations demonstrate the multistability of rotating waves, which we attribute to the dynamical system representing the state of the MSC flow tending to move along the same solution branch of the bifurcation diagram when ${\rm Ha}$ is varied. In experiments with both fixed and varying ${\rm Ha}$, the rotation frequencies of the rotating waves are consistent with the results of nonlinear stability analysis. A brief numerical investigation shows that differences in the azimuthal wavenumbers of the rotating waves that develop in the flow also depend on the azimuthal modes that are initially excited.
\end{abstract}

\maketitle


\section{\label{sec:introduction}Introduction}
The interaction of magnetic fields with the flow of electrically conducting fluids and plasmas is a common astrophysical phenomenon. The dynamo theory postulates that planetary, stellar, and galactic magnetic fields are self-excited in flows of electrically conducting fluids and plasmas. Another important effect is the magnetorotational instability (MRI), a process by which differentially rotating conducting fluids are destabilised in the presence of magnetic fields. The MRI is conceivably responsible for the outward angular momentum transport in accretion disks \cite{Balbus1991}, which surround stars and black holes. Experimentally, the dynamo theory has been successfully demonstrated in Riga \cite{Gailitis2000}, Karlsruhe \cite{Stieglitz2001}, and Cadarache \cite{Monchaux2007}. Experimental evidence of the MRI has been found at Helmholtz-Zentrum Dresden-Rossendorf in Dresden, for both helical \cite{Stefani2006} and azimuthal \cite{Seilmayer2014} geometries of the applied magnetic field. However, despite great efforts and promising initial results obtained in spherical Couette \cite{Sisan2004} and Taylor-Couette \cite{Nornberg2010} experiments, as well as in an interesting spring-mass analogue \cite{Hung2019}, the unequivocal proof of the standard MRI, with a purely axial geometry of the applied magnetic field, has not yet been found.

A spherical Couette flow is induced in a fluid filling the void between two concentric spheres when either or both of the spheres are rotated so that the fluid rotates differentially. In the most common case, the inner sphere is rotated while the outer sphere remains stationary. Such a flow is fully defined by the ratio of the radii of the inner to the outer spheres $r_{\rm i}/r_{\rm o}$ and by the Reynolds number ${\rm Re} = \Omega\,{r_{\rm i}}^2 / \nu$, where $\Omega$ and $\nu$ are the rotation speed of the inner sphere and the kinematic viscosity of the fluid, respectively. If the fluid is electrically conducting and a magnetic field is applied, the flow is said to be `magnetised'. Thus, the hydrodynamical problem of the spherical Couette flow becomes a magnetohydrodynamical one. This `magnetisation' results in the \textit{magnetic spherical Couette} (MSC) flow, which requires the magnetic Prandtl number ${\rm{Pm} = \nu / \eta}$ (where $\eta$ is the magnetic diffusivity of the fluid) and the Hartmann number ${{\rm Ha} = B_0r_{\rm i} / {{\mu}_0 \rho \nu \eta}^{1/2}}$ (where $B_0$, ${\mu}_0$, and $\rho$ are the applied magnetic field, vacuum permeability, and the density of the fluid, respectively) to be considered as additional governing parameters. The MSC flow displays different features depending on the geometry of the applied magnetic field and the electrical conductivity of the spheres. Detailed explanations of these differences were discussed by Hollerbach \cite{Hollerbach2009} and Gissinger \cite{Gissinger2011}. The base state of the spherical Couette flow is an axisymmetric flow with the primary flow driven directly by the rotation of the inner sphere. As the fluid rotates with the inner sphere, it is driven outward by inertia, thus creating a secondary meridional circulation consisting of a narrow jet towards the outer sphere along the equator, and a return flow from the rest of the outer sphere towards the origin of the jet\cite{Hollerbach2009}. Spherical Couette flows typically encounter three types of instabilities --- referred to as radial jet, return flow, and shear layer --- depending on whether the kinetic energy of the instabilities are concentrated in the jet or in the return flow of the meridional circulation, or in the shear layers along an imaginary cylinder tangent to the inner sphere that runs parallel to the axis of rotation (the so-called tangent cylinder). The precise regimes where these instabilities occur and their descriptions are discussed in Sec.\,\ref{sec:equations}.

The MSC experiment shares similarities with both the dynamo and the MRI experiments. Due to its geometry, the MSC flow models spherical cosmic dynamos such as the liquid iron within the Earth, the hydrogen plasma within the Sun, and the metallic hydrogen within Jupiter. Indeed, the MSC flow exhibits dynamo instability when the magnetic Reynolds number $\rm{Rm}$~=~${\rm{Re}\,\nu / \eta}$~=~$\rm{Re}\,\rm{Pm}$ is in the order of magnitude of a few thousands \cite{Sisan2004} and ${\rm Pm}$ is greater than 1\cite{Guervilly2010}. The fundamental feature of the MSC flow, that is, the exposure of differentially rotating, electrically conducting fluids to magnetic fields, is the same as the condition required for the onset of the MRI. Indeed, Sisan \textit{et al.}\cite{Sisan2004} observed  spontaneous excitation of oscillating magnetic and velocity field pertubations in their MSC experiment in Maryland, which have been attributed to the MRI. Reinforcing their claim was the observation of increased applied torque accompanying the fluctuations, indicating an increased transport of angular momentum from the inner sphere to the outer sphere --- a key signature of the MRI. Later on, however, Hollerbach\cite{Hollerbach2009} numerically investigated the MSC flow with the same configuration as the experiments conducted by Sisan \textit{et al.}, but at much lower values of ${\rm Re}$ (between 250 and 3000, as opposed to between $1.7 \times 10^5$ and $3.4 \times 10^6$). Hollerbach observed the radial jet and the return flow instabilities and suggested that the results of Sisan \textit{et al.} may have been turbulent analogues of these instabilities. Gissinger \textit{et al.}\cite{Gissinger2011} numerically investigated the MSC flow with a similar configuration as by Sisan \textit{et al.} Despite observing several similarities between their results, Gissinger \textit{et al.} disputed the claim that the instabilities observed by Sisan \textit{et al.} were the MRI. The major concern was that induction is a prerequisite for the generation of standard MRI, whereas the instabilities found by Gissinger \textit{et al.} were basically inductionless. For this reason, and judging from the other similarities in both results, Gissinger \textit{et al.} concluded that the instabilities observed by Sisan \textit{et al.} were probably the return flow and the shear layer instabilities. Furthermore, Gissinger \textit{et al.} observed that both the return flow and the shear layer instabilities are capable of efficiently transporting angular momentum outwards, thus proving that angular momentum transport due to magnetohydrodynamic instabilities is not restricted to, and hence, not a sufficient proof of, the existence of the MRI.

Apart from the setup in Maryland where the experiments by Sisan \textit{et al.}\cite{Sisan2004} were conducted, the Derviche Tourneur Sodium (DTS) in Grenoble is another MSC setup that has been used prominently for several MSC experiments\cite{Schmitt2008, Nataf2008, Brito2011, Schmitt2013, Figueroa2013, Nataf2013, Cabanes2014}. The setups in Maryland and Grenoble both use liquid sodium as the working fluid, but unlike in Maryland, where an axial magnetic field is applied to the flow, the DTS utilises a dipolar magnetic field of a magnet inside the inner sphere. 

In this work, we present results obtained with the HEDGEHOG (Hydromagnetic Experiment with Differentially Gyrating sphEres HOlding GaInSn) MSC setup at Helmholtz-Zentrum Dresden-Rossendorf. As the name implies, the working fluid of HEDGEHOG is Ga\textsuperscript{67}In\textsuperscript{20.5}Sn\textsuperscript{12.5}, a liquid metal eutectic alloy. Apart from the difference in the working fluid, HEDGEHOG differs from the experimental setups in Maryland and Grenoble in terms of the conductivity of the spheres. Whereas the other two setups consist of electrically conducting inner spheres and insulating outer spheres, both spheres in HEDGEHOG are insulating. HEDGEHOG employs an axial magnetic field geometry like the setup in Maryland. In a first set of experiments with HEDGEHOG, published by Kasprzyk \textit{et al.} \cite{Kasprzyk2017}, the radial jet instability, the lower ${\rm Ha}$ range of the return flow instability, and the stable transition region between them were investigated. Our aim is to investigate the return flow instability in much more detail. 

MSC instabilities directly manifest from the base state as waves travelling in the azimuthal direction with constant rotation frequency, which are known as rotating waves. Golubitsky \textit{et al.} \cite{Golubitsky2000} documented the occurrence of rotating waves in several experiments. They cited rotating waves appearing as wavy vortices in Taylor-Couette systems \cite{Andereck1986}, as spiral waves in the Belousov-Zhabotinsky chemical reaction \cite{Winfree1980, Li1996, Skinner1991} and Rayleigh-Bernard convection \cite{Plapp1996}, and as cellular patterns in laminar premixed flames \cite{Gorman1994}. Rotating waves have also been observed in non-magnetic spherical Couette flows. Wulf \textit{et al.} \cite{Wulf1999} experimentally investigated the path followed by the base state of the non-magnetic spherical Couette flow as it underwent two Hopf bifurcations on its way to chaos, the first of which resulted in the appearance of rotating waves. In light of the computation and nonlinear stability analysis of the rotating waves in an MSC flow conducted by Garcia and Stefani \cite{Garcia2018} at conditions and configuration corresponding to HEDGEHOG, we compare their numerical results with our experimental results in a regime where rotating waves with different wavenumbers are stable. The present study focuses on the comparison of the rotation frequencies of the rotating waves. We conduct the experiments at both fixed and varying values of ${\rm Ha}$. The latter approach helps to investigate the multistability of rotating waves through the trajectories of the MSC flow state as a dynamical system in state space. The multistability of rotating waves implies that at certain values of ${\rm Ha}$, a rotating wave can assume one of several possibile wavenumbers.


\section{\label{sec:equations}Equations, numerical model, and instability regimes}
If the spheres in the MSC flow rotate about ${\mathbf{\hat{e}}_z}$, the unit vector representing the $z$-axis, the axial magnetic field applied externally to the flow is given by
\begin{equation}
	\mathbf{B}_0 = B_0\cos{(\theta)}\hat{e}_r - B_0\sin{(\theta)}\hat{e}_\theta
\end{equation}
where $\theta$, $\hat{e}_r$, and $\hat{e}_\theta$ are the magnetic field, the colatitude, and the unit vectors representing the radial distance and the colatitude, respectively. The fluid flow $\mathbf{v}$ is governed by the Navier-Stokes equation
\begin{equation}
	\partial_t\mathbf{v} + \mathbf{v} \cdot  \nabla v = - \nabla p + {\rm Re}^{-1} \nabla^2 \mathbf{v} + {\rm Ha}^2 {\rm Re^{-1}} (\nabla \times \mathbf{b}) \times \mathbf{\hat{e}}_z
\end{equation}
where $\mathbf{b}$ is the deviation of the magnetic field induced by the interaction of the fluid flow with $\mathbf{B}_0$. For ${\rm Rm} << 1$, this induced magnetic field is negligible compared to $\mathbf{B}_0$. The so-called inductionless approximation \cite{Mueller2001} applies here, since all the experiments were conducted at ${\rm Re}$~=~1000 and the ${\rm Pm}$ of GaInSn is in the order of $10^{-6}$, giving rise to ${\rm Rm}$ in the order of $10^{-3}$. The induction equation simplifies to
\begin{equation}
	0 = {\nabla}^2 \mathbf{b}+\nabla \times (\mathbf{v} \times \mathbf{\hat{e}}_z).
\end{equation}

Garcia and Stefani \cite{Garcia2018} discretised and integrated the equations using a pseudo-spectral method described in Garcia \textit{et al.} \cite{Garcia2010} to obtain a numerical model corresponding to experiments with HEDGEHOG.

\begin{figure}
\includegraphics[width=\columnwidth]{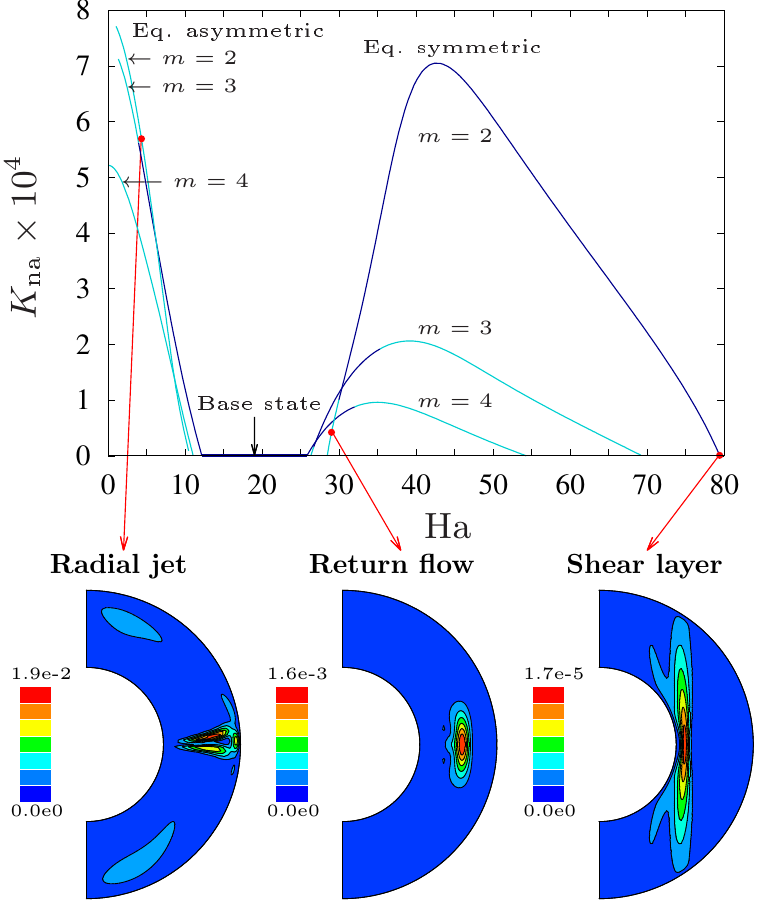}
	\caption{\label{fig:bifurcation}The non-axisymmetric kinetic energies in a spherical Couette flow for $r_{\rm i} / r_{\rm o}$~=~0.5 and ${\rm Re}$~=~1000 shown in a bifurcation diagram for ${\rm Ha} \in [0,80]$ (top) and contour plots at different instability regimes (bottom). In the bifurcation diagram, $K_{\rm na}$ denotes the non-axisymmetric kinetic energy. The set of lines to the left and right of the base state represent equatorially asymmetric and symmetric kinetic energies, respectively. The dark blue and light blue lines represent the kinetic energies of stable and unstable rotating waves, respectively.}
\end{figure}

For any given $r_{\rm i} / r_{\rm o}$, the parameter space of the spherical Couette flow governed by ${\rm Re}$ and ${\rm Ha}$ can be partitioned by stability curves of critical ${\rm Re}$ separating regions of stability from those of instability. Such curves were computed using linear stability analysis by Hollerbach\cite{Hollerbach2009} and Travnikov \textit{et al.}\cite{Travnikov2011} The stable region of the spherical Couette flow is the axisymmetric base state. Crossing the stability curves from the base state introduces non-axisymmetric instabilities in the flow. The instabilities manifest as rotating waves whose azimuthal wavenumbers $m$ depend on the stability curve that is crossed. The rotating waves result from Hopf bifurcations from the base state. In the bifurcation diagram of the dynamical system representing the state of the flow, every point on the branches is a solution of the rotating wave given by $m$. Bifurcation diagrams may be obtained for several time and volume-averaged properties such as the kinetic energies of the flow and the rotation frequencies of the rotating waves\cite{Garcia2018}. The bifurcation diagram of the non-axisymmetric kinetic energy computed by means of Newton-Krylov continuation techniques for periodic orbits \cite{Sanchez2016} for ${r_{\rm i} / r_{\rm o} = 0.5}$, ${\rm Re}$~=~1000, and ${\rm Ha} \in [0,80]$, is shown at the top of Fig.\,\ref{fig:bifurcation}. 

The bifurcation diagram precisely shows the stability regions of rotating waves with $m$~=~2, 3, and 4. In the base state, the flow is stable and no rotating wave exists. Hence, this state possesses no kinetic energy in the non-axisymmetric azimuthal modes. It is driven by the rotation of the inner sphere, which flings the fluid outwards by inertia, thus creating a secondary meridional circulation consisting of a narrow jet along the equator towards the outer sphere, and a return flow from the rest of the outer sphere towards the origin of the jet \cite{Hollerbach2009}. The region with finite values of non-axisymmetric kinetic energy to the left of the base state in the bifurcation diagram is characterised by an equatorially asymmetric instability known as the radial jet instability. The non-axisymmetric kinetic energy of the radial jet instability, as seen in the contour plots at the bottom left of Fig.\,\ref{fig:bifurcation}, is concentrated on the jet along the equator in the meridional circulation. The jet adopts a wavy structure, alternatively slightly above and below the equatorial plane \cite{Hollerbach2009}. We also see in the bifurcation diagram that the radial jet instability occurs even when no magnetic field is applied. 

Increasing the magnetic field from the radial jet instability regime, while maintaining the rotation of the inner sphere, re-stabilises the flow. The next regime, to the right of the base state in the bifurcation diagram, consists of equatorially symmetric instabilities. We see that the non-axisymmetric kinetic energies of the rotating waves with each $m$ continuously increase until they attain their respective maxima, and then continuously decrease. At the lower ${\rm Ha}$ range of the equatorially symmetric instabilities, that is, where the non-axisymmetric kinetic energies increase, is the return flow instability. The kinetic energy of this instability is concentrated on the return flow portion of the meridional circulation, as seen in the contour plot at the centre. At the higher ${\rm Ha}$ range of the equatorially symmetric instabilities, where the non-axisymmetric kinetic energies decrease, is the shear layer instability. As the spherical Couette flow becomes increasingly `magnetised', shear layers begin to concentrate along the tangent cylinder. We see in the contour plot on the right that the kinetic energy of the shear layer instability is concentrated in these shear layers. The feature of the instability consists of series of vortices in the horizontal flow spawned by shear layers at the outer edge\cite{Hollerbach2001}.

\section{Experimental methods}
\subsection{\label{sec:apparatus}Apparatus}
\begin{figure}
	\includegraphics[width=\linewidth]{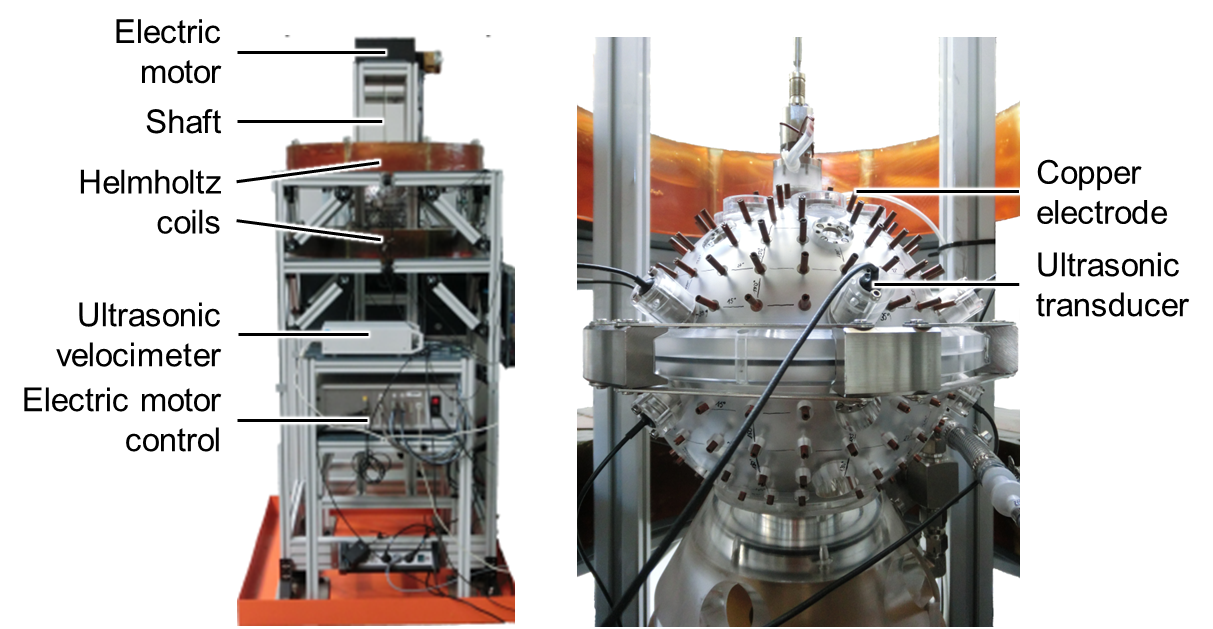}
        \caption{\label{fig:hedgehog}Photographs of the HEDGEHOG setup showing its full view (left) and a close-up view on the outer sphere (right).}
\end{figure}

Photographs of the HEDGEHOG setup are shown in Fig.\,\ref{fig:hedgehog}. The radii of the inner and outer spheres are 45 mm and 90 mm, respectively. The spheres are made of acrylic glass. The inner sphere is rotated by an electric motor shaft passing through the spheres along their common axis. The diameter of the shaft is very small (3 mm) relative to the fluid volume, so that its interference with the flow is minimal. A lead weight within the inner sphere counterbalances the significant amount of buoyancy exerted by GaInSn owing to the high density of the fluid. A Helmholtz-type coil pair, consisting of 80 wire windings in each coil, generates a nearly uniform axial magnetic field. By definition, the radii of the coils and the distance between them in a Helmholtz coil are all equal. However, this condition is only approximately satisfied in HEDGEHOG, as the radius of each coil is 300 mm, while the distance between them is 310 mm. The outer sphere has five latitudinal rows of 6 azimuthally equidistant provisions for mounting ultrasonic transducers for flow velocity measurement. Additionally, the outer sphere is fitted with 168 needle-like copper electrodes, providing another justification for naming the experimental setup `HEDGEHOG'. These electrodes were provided for experiments necessitating electric potential measurements.

\subsection{\label{sec:instrumentation}Instrumentation}
\begin{figure}
\includegraphics[height= 5 cm, keepaspectratio]{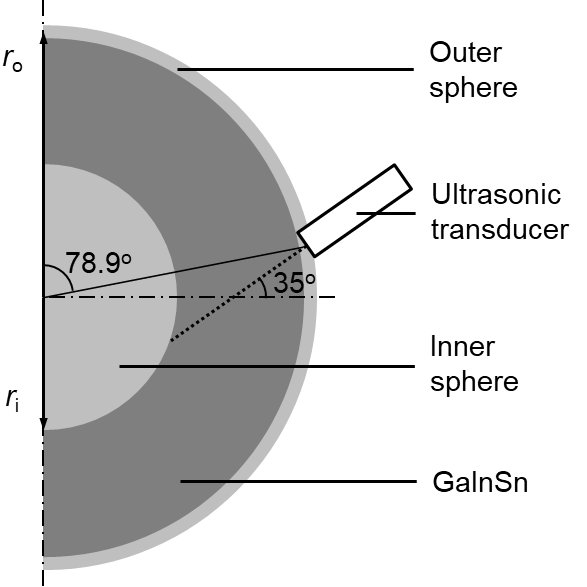}
	\caption{\label{fig:transducer}Meridional cross-section through the axis of one of the 6 azimuthally equidistant ultrasonic transducers installed on HEDGEHOG for the experiments. The dotted line extending from the ultrasonic transducer represents its line of view along which the flow velocity components parallel to the line were measured.}
\end{figure}

Ultrasonic Doppler Velocimetry was employed for the flow velocity measurement. This technique utilises piezoelectric ultrasonic transducers (Signal Processing SA, model TR0410LS) and an ultrasonic Doppler velocimeter (Signal Processing SA, model DOP3010). For our experiments, 6 transducers were installed at a colatitude of 78.9\degree. Each transducer was inclined southwards so that the angle between the axis of each transducer and the equator was 35\degree. Fig.\,\ref{fig:transducer} shows the meridional cross-section through one of the transducers.  

An outer sphere wall thickness of 7 mm separates the measuring surface of the transducer and the GaInSn. An ultrasonic gel was applied between the transducers and the outer sphere to ensure good coupling between the surfaces and facilitate the paths for the ultrasonic pulses into the GaInSn. Each transducer periodically emitted an ultrasonic pulse at a frequency of 4 MHz and received the echoes from the particles in the path of the ultrasonic beam. Each transducer was connected to an individual channel on the velocimeter. From each channel, the velocimeter sequentially acquired the echo signals reflected by particles in the fluid to determine the distances between the particles and the transducer using the velocity of sound in the GaInSn. GaInSn oxidises readily in air \cite{Shafiei2015}, which produces solid particles within it, thus eliminating the need to artificially suspend particles in the fluid \cite{Borin2009}. The velocities of the particles were determined from their distances after successive pulse emissions. Using this information, the velocity signals of the flow over distances parallel to each transducer were inferred.


\subsection{\label{sec:procedure}Procedure}
Before each experimental run, the GaInSn was mixed to distribute the particles within the fluid for proper velocity measurement. For this purpose, a peristaltic pump was attached to a tube externally connecting the poles of the outer sphere. The pump continuously circulated the GaInSn from the south pole to the north pole, thereby distributing the particles in the fluid. The particles remained sufficiently distributed for about eight hours, beyond which too much of the particles would have sedimented for realistic velocity measurement. Hence, the length of the experiment was limited by the sedimentation time of the particles. 

The magnetic field $B_0$ corresponding to the desired ${\rm Ha}$ was produced by the current $I$ supplied to the Helmholtz coil. The relationship between $B_0$ and $I$ is given by 
\begin{equation}
	B_0 = \left(\frac{4}{5}\right)^{3/2} \frac{\mu_0 n I}{R}
\end{equation}
where $n$ is the number of wire windings in each Helmholtz coil and $R$ is the radius of each coil and the distance between the coils. As mentioned in Sec.\,\ref{sec:apparatus}, the distance between the coils and their radii are only approximately equal in HEDGEHOG. The distance between the coils (310 mm) was used as $R$ to evaluate $B_0$. The desired ${\rm Re}$ was set by the rotation of the inner sphere by the electric motor. The physical properties of GaInSn, listed in Morley \textit{et al.} \cite{Morley2008}, were approximated to two significant figures and used to evaluate all the relevant parameters. The values of these properties are listed in Table \ref{tab:properties}.
\begin{table}
\caption{\label{tab:properties}Physical properties of Ga\textsuperscript{67}In\textsuperscript{20.5}Sn\textsuperscript{12.5} from Morley \textit{et al.} \cite{Morley2008} corrected to two significant figures.}
\renewcommand{\arraystretch}{1.2}
\begin{ruledtabular}
\begin{tabular}{c c c}
Density & Kinematic viscosity & Electrical conductivity \footnote{$\mu_{\rm 0}$ and $\eta$ are vacuum permeability and magnetic diffusivity, respectively.} \\
	$\rho\;(\rm kg\;m^{-3})$ & $\nu\;(\rm m^2\;s^{-1})$ & $\sigma = 1/\mu_{\rm 0}\,\eta\;(\rm {\Omega}^{-1}\;m^{-1})$ \\
\hline
$6.4 \times 10^3$ & $3.0 \times 10^{-7}$ & $3.1 \times 10^6$\\
\end{tabular}
\end{ruledtabular}
\end{table}


\section{Results and discussion}
\subsection{\label{sec:testAndData}Test conditions and data analyses}
The experimental investigation followed two approaches. In the first approach, each experiment was conducted at fixed ${\rm Ha}$, while in the second approach, ${\rm Ha}$ was varied over the duration of each experiment. The first approach enables the flow to attain a fully saturated state. Meanwhile, the second approach offers good insight into the trajectory of the dynamical system in state space. We conducted all the experiments at ${\rm Re} = 1000$ to correspond with the conditions in the numerical investigation by Garcia and Stefani \cite{Garcia2018}. At ${\rm Re}$~=~1000, Travnikov \textit{et al.} \cite{Travnikov2011} computed ${\rm Ha}$~=~25.8 as the critical ${\rm Ha}$, above which the flow becomes unstable to the return flow instability. At this critical ${\rm Ha}$, the flow is destabilised so that a rotating wave with $m$~=~4 develops in the flow (see the bifurcation diagram in Fig.\,\ref{fig:bifurcation}). To ensure that we are in the return flow instability regime, we selected the lower bound of the ${\rm Ha}$ for the experiments to be slightly above this critical value, at ${\rm Ha}$~=~27.5. The instability curves and the branches of the bifurcation diagram for the return flow and the shear layer instabilities are continuously connected\cite{Hollerbach2009, Garcia2018}, which means that the upper bound of ${\rm Ha}$ for the return flow instability, where it transitions to the shear layer instability, is not clearly defined. The transition from the return flow to the shear layer instabilities does not occur abruptly, but the flow contains features of both instabilities during the transition process\cite{Garcia2018}. We have seen how the non-axisymmetric kinetic energies of the flow for each $m$ continuously increase, and then decrease after attaining their maxima in the bifurcation diagram of Fig.\,\ref{fig:bifurcation}. Garcia and Stefani\cite{Garcia2018} suggested that the points at which the non-axisymmetric kinetic energies reach their maximum values could be defined as the transition points between the return flow and the shear layer instabilities. These maxima are different for each of $m \in \{2,3,4\}$. Of these, the maximum for $m = 2$ occur at the highest value of ${\rm Ha}$. Furthermore, only the solution branch of $m = 2$ represents that of a stable rotating wave at the maximum value of its non-axisymmetric energy, whereas the rotating waves with $m$~=~3 and 4 become unstable before attaining their respective maxima. The $m = 2$ solution attains its maximum value of non-axisymmetric kinetic energy between ${\rm Ha}$~=~40 and 45, the former of which we select as the upper bound of the ${\rm Ha}$ for the experiments to once again ensure that the experiments are conducted within the return flow instability regime. 

Based on the considerations in Sec.\,\ref{sec:procedure} regarding the sedimentation time of GaInSn particles, the length of each experiment was chosen to be 6 hours. The flow velocity signals from all 6 ultrasonic transducers were analysed at a specific depth using a two-dimensional fast Fourier transform (FFT) in time and azimuthal positions of the transducers. The transducers did not acquire the velocity signals simultaneously, but instead, successively acquired the signals from one transducer at a time. There was a time delay of about 1.2 seconds between successive signal acquisitions. Since this time delay is about two orders of magnitude less than the rotation periods of the rotating waves in the MSC flow, the effect on the computation of the rotation frequencies of the experimentally observed rotating waves was deemed negligible. Thus, 6 successive signals, each from a different transducer, were grouped into a single time step to form the velocity-time data to be analysed using the FFT. The FFT was implemented using a short-time Fourier transform (STFT) to account for the non-stationarity of the data. The FFT was applied within several rectangular time windows of fixed length. The window was initially placed at the start of the data, and the FFT process was reiterated as the window was shifted by a constant time interval over the entire data. For our analyses, the length of each time window was 7200 seconds and the window was slid by 600 seconds after each iteration. Thus, 25 time windows were used to analyse the data for each 6-hour experiment, with the first time window spanning between 0 and 7200 seconds, and the last time window spanning between 14400 seconds and the end of the experiment at 21600 seconds. The temporal frequency resolution of the FFT was 1/7200 Hz.


\subsection{\label{sec:fixedHa}Experiments with fixed ${\rm Ha}$}
\begin{figure*}
	\includegraphics[width=\textwidth]{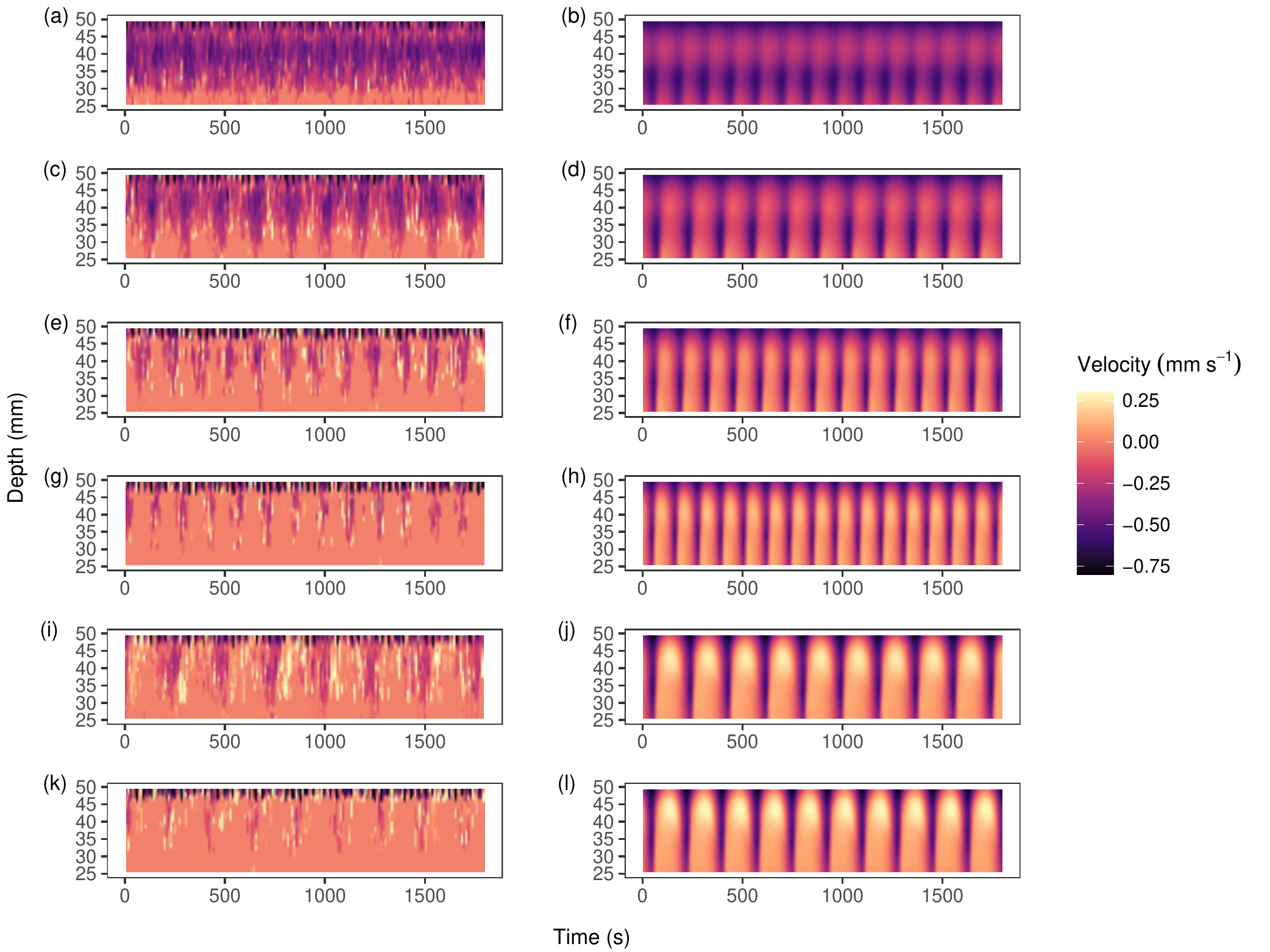}
	\caption{\label{fig:velocities}Plots of the experimentally measured (left column) and numerically simulated (right column) velocity components parallel to the ultrasonic transducer axis. The depth indicates the axial distance from the ultrasonic transducer. ${\rm Ha}$~=~(a, b) 27.5, (c, d) 30, (e, f) 32.5, (g, h) 35, (i, j) 37.5, and (k, l) 40. For the numerical results, $m$~=~(a) 4, (d, f, h) 3, and (j, l) 4. ${\rm Re}$~=~1000 in all cases.}
\end{figure*}

Figure \ref{fig:velocities} shows the plots of the velocity components parallel to the axis of an ultrasonic transducer in HEDGEHOG at different values of ${\rm Ha}$ over a period of 30 minutes. The column on the left of Fig.\,\ref{fig:velocities} shows the velocities measured during the experiments, while the column on the right shows the numerically simulated velocities. In order to ensure that the flows have attained saturated state, the velocities were obtained close to the end of each experiment. The velocities were taken along the line of view of the transducer, represented by the dotted line extending from the transducer in Fig.\,\ref{fig:transducer}. The depth increases as the distance from the transducer. The transducers were located at the outer sphere above the equator at $\theta = 78.9 \degree$, and they pointed towards the inner sphere. The sign convention is such that flows away from the transducer and those towards the transducer have positive and negative velocities, respectively. We observe that the general structure of the plots for both the measured and numerically simulated velocities consists of periods of velocities close to zero separated by periods of negative velocities, although the separations of the periods in the numerically simulated velocities is visibly clearer. As explained in Sec.\,\ref{sec:equations}, the flow feature of the spherical Couette flow consists of meridional circulation from the inner sphere towards the outer sphere along the equator (the radial jet), and back from the outer sphere towards the inner sphere over the void between the spheres (the return flow). The return flow crosses the viewing line of the transducers nearly perpendicularly (hence, it is measured as having velocities close to zero by the transducer), while the radial jet flows from the inner sphere to the outer sphere, where the transducers are located (hence, it is measured as having negative velocities by the transducer). Therefore, what we observe in the velocity plots in Fig.\,\ref{fig:velocities} are alternating cells of return flows and radial jets. The frequency of alternation corresponds to the frequency of the rotating waves in the flow. In all the velocity plots, the regions representing the radial jet are narrowest at shallower depths, close to the outer sphere. The regions of the radial jet gradually widen as the depth increases until they reach a certain depth, beyond which the regions of the radial jet begin to narrow. The depths at which the regions of the radial jet are the widest before they narrow are points where the transducer's line of view intersects with the equator of the spherical Couette sytem. This is as expected, since at any radial distance, the velocity of the radial jet is maximum along the equator. As we approach the surface of the inner sphere (located at a depth of 57 mm), the regions of the radial jet become more dominant in the velocity plots. This is again as expected, since the radial jets originate from the inner sphere. 

Due to the multistability of rotating waves, multiple solutions of the numerically simulated velocities exist at certain values of ${\rm Ha}$. Considering $m\in\{0, 1, 2, 3, 4\}$, there are solutions of $m = 2$ for ${\rm Ha} \in \{37.5,40\}$, $m = 3$ for ${\rm Ha} \in \{27.5, 30, 32.5, 35\}$, and $m = 4$ for ${\rm Ha} \in \{27.5, 30, 32.5\}$. The numerically simulated velocity plots in Fig.\,\ref{fig:velocities} were selected from solutions whose periodicities most closely match those of the experimental velocity plots. These solutions were  $m = 4$ for ${\rm Ha} = 27.5$, $m = 3$ for ${\rm Ha} \in \{30, 32.5, 35\}$, and $m = 2$ for ${\rm Ha} \in \{37.5, 40\}$. Hence, we may make an initial guess that these were the azimutal wavenumbers of the rotating waves present in the flow during the experiments conducted at the corresponding values of ${\rm Ha}$.

\begin{figure}
\includegraphics[width=\columnwidth]{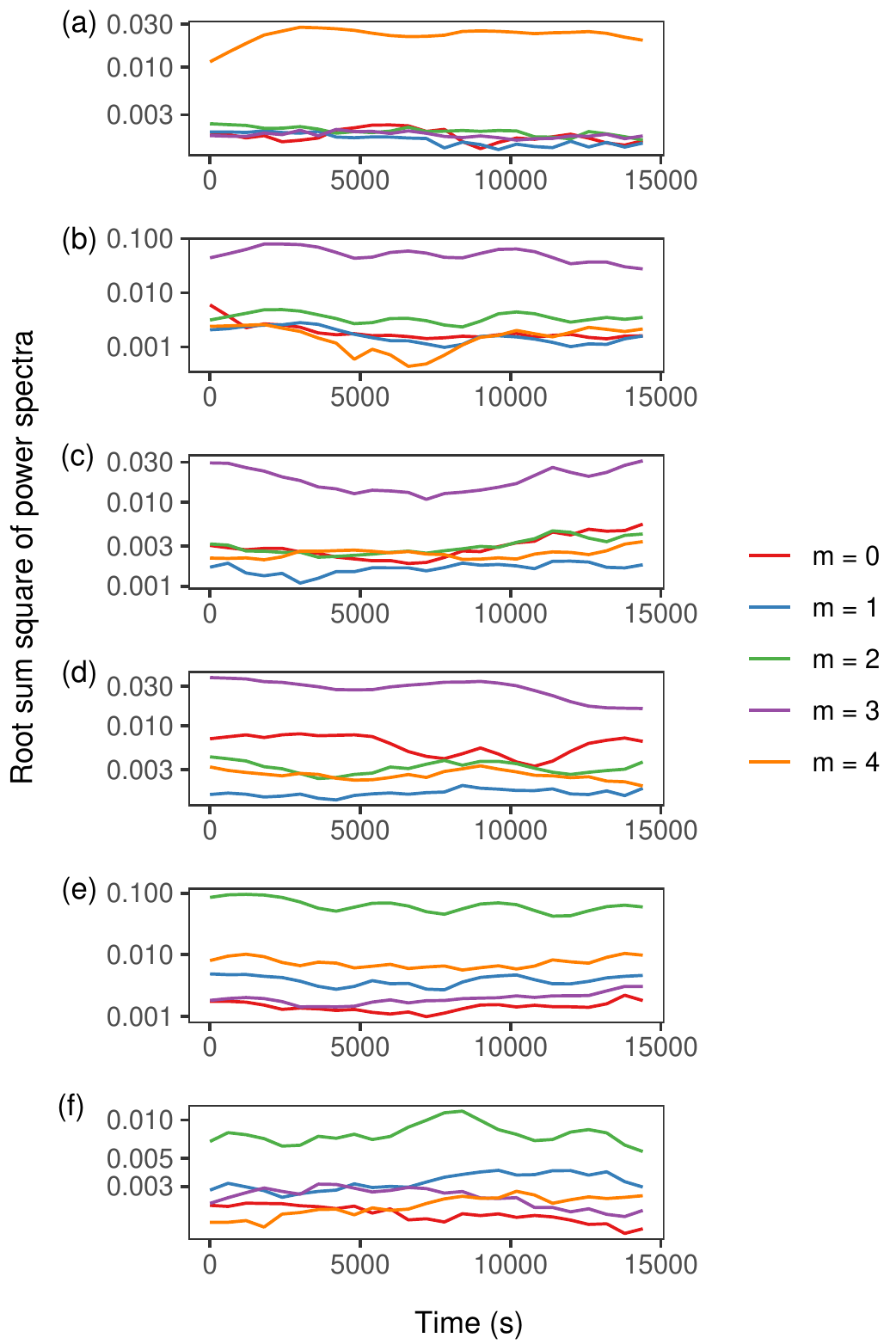}
\caption{\label{fig:energies}Root sum squares of the normalised power spectra in each azimuthal mode obtained from the STFT of the velocity-time data. ${\rm Ha}$~=~(a) 27.5, (b) 30, (c) 32.5, (d) 35, (e) 37.5, and (f) 40. ${\rm Re}$~=~1000 in all cases.}
\end{figure}

In order to quantitatively determine the azimuthal wavenumbers of the rotating waves developed in the flow during the experiments, we computed the power spectra of the velocity-time data. The absolute values of the FFT amplitudes were squared, and the results were divided by twice the variance of the velocities in each STFT time window to obtain the normalised power spectra. Finally, the root sum square (RSS) of the power spectra in the azimuthal modes were compared. The depths at which the STFT were implemented was important, since as we see in Fig.\,\ref{fig:velocities}, the experimental velocity plots were not well-defined in terms of the separation between the radial jets and the return flows, especially at shallower depths. Depths with high ratios of the largest RSS of the power spectra to the RSS of the power spectra belonging to the rest of the azimuthal modes were selected for the implementation of the STFT. These depths, according to Fig.\,\ref{fig:velocities}, were 30~mm for ${\rm Ha}$~=~27.5 and 45~mm for ${\rm Ha \in\{30,32.5,35,37.5,40\}}$. Depths of 30 mm and 45 mm correspond to ($r$, $\theta$)~=~(64 mm, 90\degree) and (53 mm, 99\degree), respectively. The RSS of the power spectra over the duration of the experiments for each ${\rm Ha}$ are shown in Fig.\,\ref{fig:energies}. The times indicate the beginning of each STFT time window. In each plot, the azimuthal mode with the largest RSS of the power spectra is evident. The plots also corroborate our guesses of the azimuthal wavenumbers of the rotating waves in the flow based on the matching periodicities between the experimental and numerical results in Fig.\,\ref{fig:velocities}.

\begin{figure}
\includegraphics[width=\columnwidth]{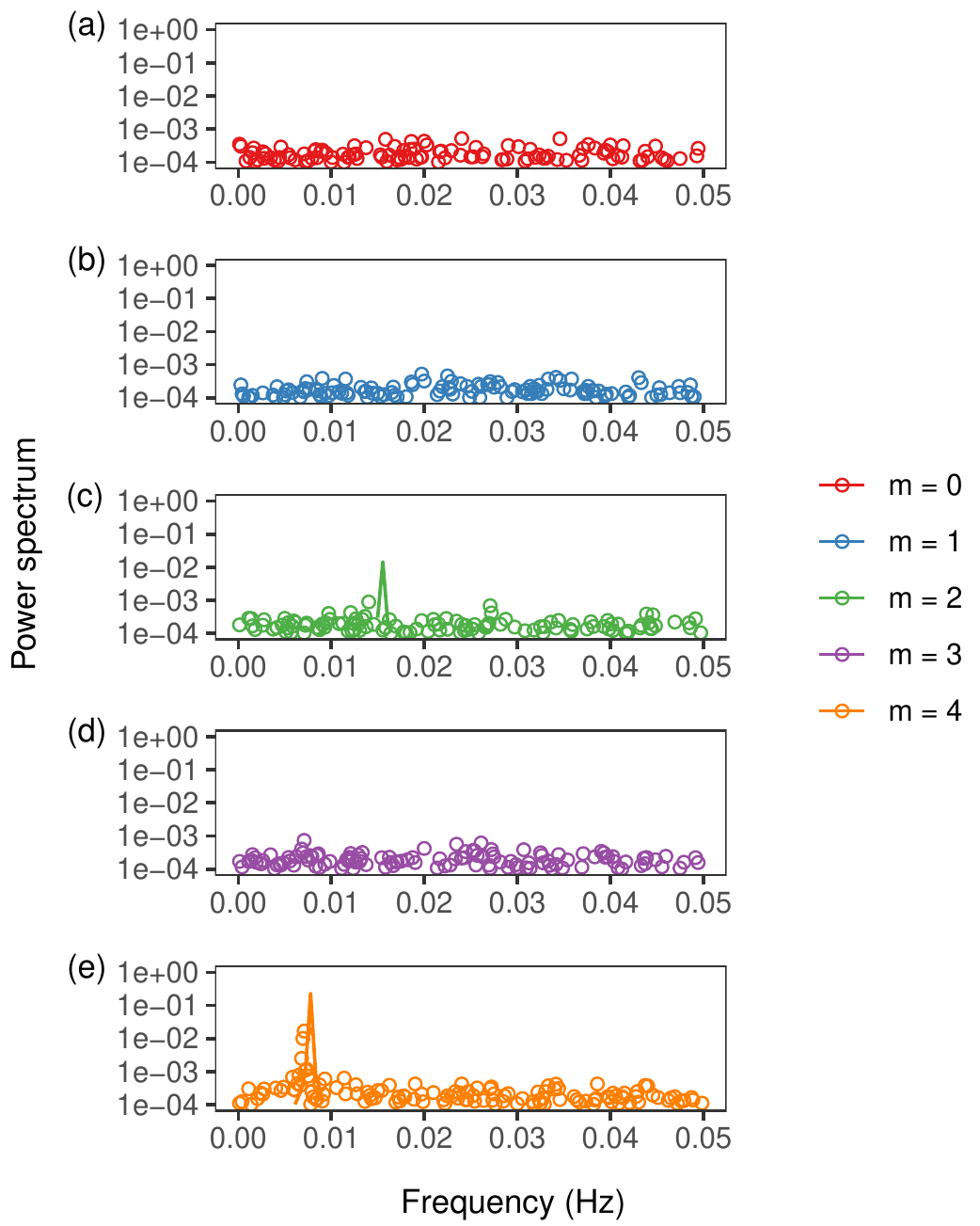}
	\caption{\label{fig:frequencies}Normalised power spectra obtained from the FFT of the velocity signal for ${\rm Ha}$~=~27.5 at $m$~=~(a)~0, (b)~1, (c)~2, (d)~3, (e)~4. The FFT was applied over a rectangular time window between 14400 and 21600 seconds in the experiment. The circles and lines represent the experimental and numerically simulated data, respectively. ${\rm Re}$~=~1000 in all cases.}
\end{figure}

We now  examine the rotation frequencies of the rotating waves, which are represented by the temporal frequencies obtained from the FFT of the velocity signals. The dimensional and dimensionless frequencies of the rotating waves are related by the equation
\begin{equation}
	f = \frac{m \omega \nu}{2\pi{(r_{\rm o} - r_{\rm i})}^2}
\end{equation}
where $f$ and $\omega$ are the dimensional frequency in Hz and dimensionless frequency, respectively. Figure \ref{fig:frequencies} shows the power spectra in each azimuthal mode obtained from the FFT of the velocity-time data for ${\rm Ha}$~=~27.5 at $m \in \{0, 1, 2, 3, 4\}$. The FFT was applied over the time window between 14400 and 21600 seconds. The data points representing the power spectra for the experimental data concentrate along the largest peak in the power spectra of the numerical data in Fig.\,\ref{fig:frequencies}(e), at $m$~=~4. Thus, a rotating wave with $m$~=~4 was present in the flow. There is a smaller peak in the power spectrum of the numerical data at $m$~=~2 in Fig.\,\ref{fig:frequencies}(c). Based on the fact that this peak occurs at a frequency that is twice the frequency of the peak at $m$~=~4, we deduced that this secondary peak is the projection of the rotating wave with $m$~=~8 in the flow.

\begin{figure}
\includegraphics[width=\columnwidth]{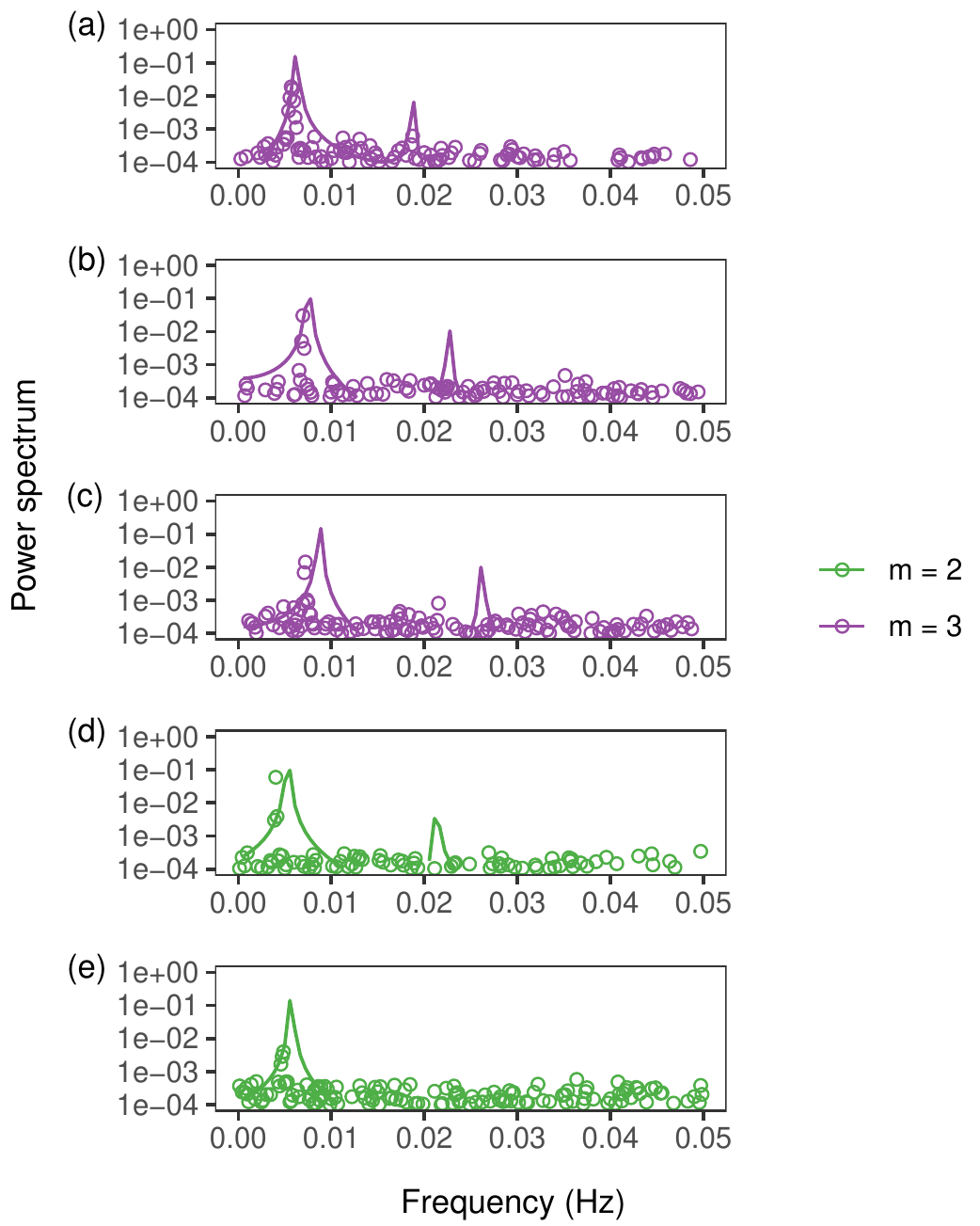}
	\caption{\label{fig:frequenciesOtherModes}Normalised power spectra obtained from the FFT of the velocity-time data for ${\rm Ha}$~=~(a)~30, (b)~32.5, (c)~35, (d)~37.5, (e)~40. The FFT was applied over a rectangular time window between 14400 and 21600 seconds in each experiment. The circles and lines represent the experimental and numerically simulated data, respectively. ${\rm Re}$~=~1000 in all cases.}
\end{figure}

Fig.\,\ref{fig:frequenciesOtherModes} shows the power spectra obtained from the FFT of the velocity-time data for ${\rm Ha} \in \{30, 32.5, 35, 37.5, 40\}$. The FFT was applied over the time window between 14400 and 21600 seconds in each case. For each experiment, one azimuthal mode had a peak in the power spectrum. These azimuthal modes, as can be seen in the figure, correspond to $m=3$ for ${\rm Ha} \in \{30, 32.5,35\}$ and $m = 2$ for ${\rm Ha} \in \{37.5,40\}$. In the numerical data, there are 2 peaks in the power spectra in all cases except for ${\rm Ha} = 40$. The smaller peaks for ${\rm Ha} \in \{30, 32.5,35\}$ occur at temporal frequencies that are each 3 times the frequencies of the larger peaks. Hence, the smaller peaks are projections of rotating waves with $m=9$ in the flow. For ${\rm Ha}=37.5$, the smaller peak occurs at a temporal frequency that is 4 times the frequency of the larger peak, implying that it is a projection of a rotating wave with $m~=~8$ in the flow. In each case, the peak in the power spectra of the experimental data concentrate on the larger peak (or on the only peak for ${\rm Ha} = 40$) of the numerical data. Therefore, the azimuthal wavenumbers of the rotating waves in the flows are 3 for ${\rm Ha} \in \{30, 32.5, 35\}$ and 2 for ${\rm Ha} \in \{37.5, 40\}$.

\begin{figure}
        \includegraphics[width=\columnwidth]{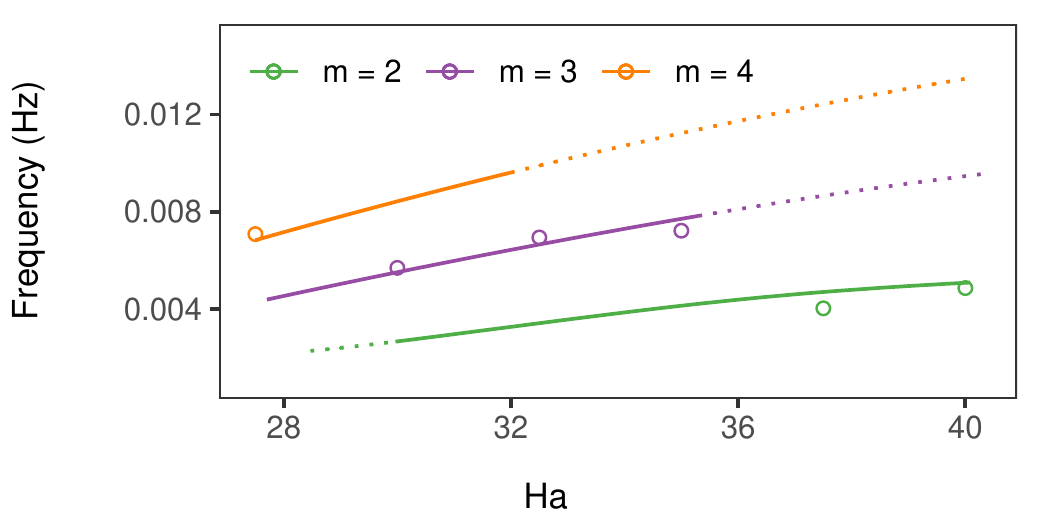}
	\caption{\label{fig:fixedHa}Rotation frequencies of the rotating waves in the flow for ${\rm Ha} \in [27.5,40]$. The circles represent the rotation frequencies obtained from the experiments with fixed ${\rm Ha}$. The solid and dotted lines represent numerical solutions of stable and unstable rotating waves, respectively. ${\rm Re} = 1000$ in all cases.}
\end{figure}

We define the rotation frequency of the rotating wave developed in the flow during the experiments as the temporal frequency with the largest power spectrum during the final time windows of the STFT (between 14400 and 21600 seconds as in Fig.\,\ref{fig:frequencies} and Fig.\,\ref{fig:frequenciesOtherModes}). This was to ensure that the rotation frequencies were evaluated after the flow had attained saturated states. Fig.\,\ref{fig:fixedHa} compares the rotation frequencies of the experimental and numerical results. We see in the figure that in general, the rotation frequencies of both results are close. We notice the most deviations in the results for ${\rm Ha} \in \{35,37.5\}$, where the rotation frequencies from the experiments were lower than the numerically computed rotation frequencies. Indeed, these deviations are observed in the velocity plots in Fig.\,\ref{fig:velocities}, where the number of cells of return flows and radial jets in the experimental data are fewer than those of the numerically simulated data for ${\rm Ha} \in \{35,37.5\}$. We attribute these differences to uncertainties such as in the physical properties of GaInSn used to evaluate ${\rm Ha}$, ${\rm Re}$, and $f$, and uncertainties in determining ${B}_0$.

\subsection{\label{sec:varyingHa}Experiments with varying ${\rm Ha}$}
\begin{figure}
\includegraphics[width=\columnwidth]{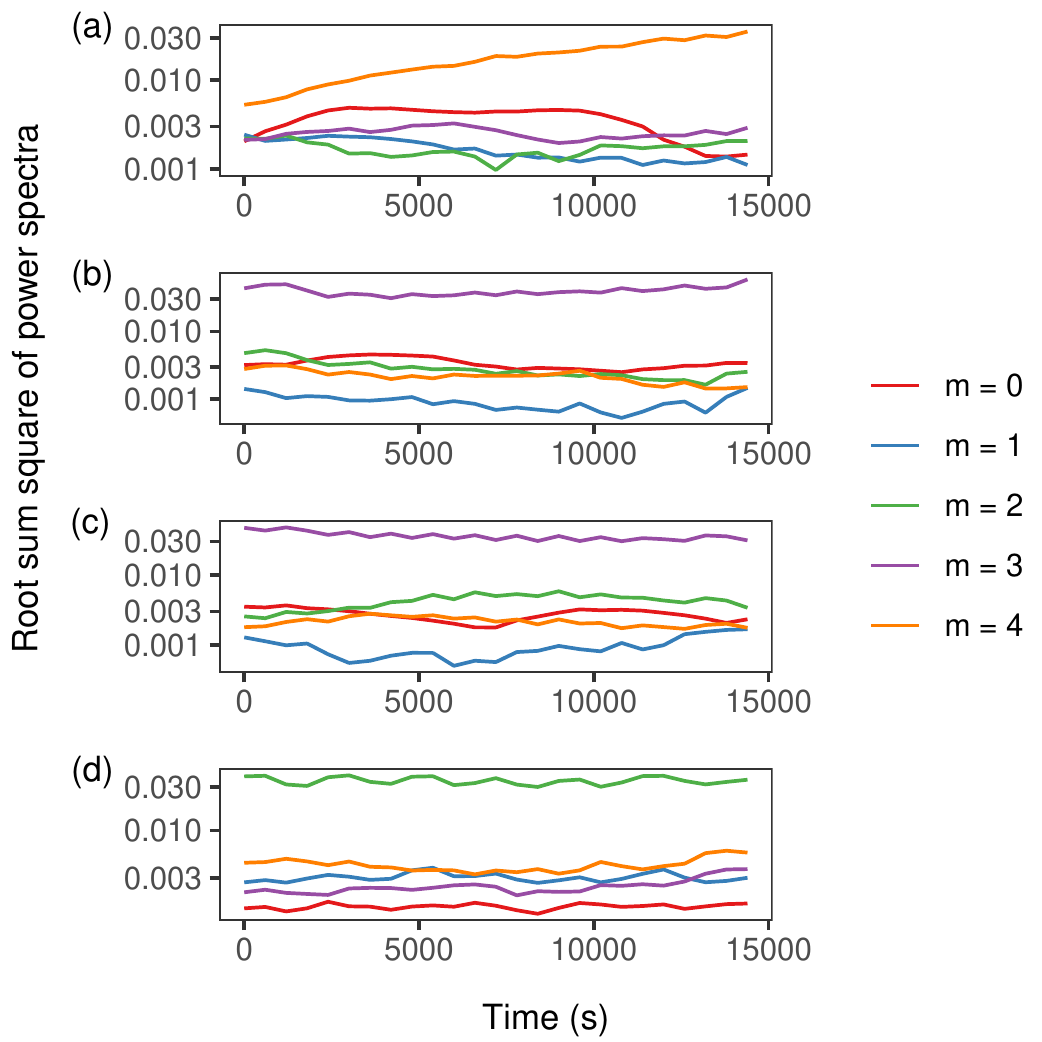}
	\caption{\label{fig:energiesVaryingHa}Root sum squares of the normalised power spectra in each azimuthal mode obtained from the STFT of the velocity-time data as ${\rm Ha}$ was (a) increased from 27.5 to 32.5, (b) increased from 30 to 35, (c) decreased from 35 to 30, and (d) decreased from 40 to 35. ${\rm Re}$~=~1000 in all cases.}
\end{figure}

In this section, we discuss the effect of linearly varying ${\rm Ha}$ between regimes where rotating waves with different azimuthal wavenumbers evolved based on results from Sec.\,\ref{sec:fixedHa}. Additionally, we investigate the potential occurence of hysteresis when ${\rm Ha}$ is increased and decreased within values where rotating waves with the same azimuthal wavenumbers were observed in experiments with fixed ${\rm Ha}$ in Sec. \ref{sec:fixedHa}. We conducted four separate experiments. In the first one, ${\rm Ha}$ was increased from 27.5 to 32.5, while in the second and third experiments, ${\rm Ha}$ was respectively increased and decreased between ${\rm Ha}$~=~30 and 35. In the last experiment, ${\rm Ha}$ was decreased from 40 to 35. In Fig.\,\ref{fig:energiesVaryingHa}, the RSS of the power spectra in each azimuthal mode obtained from the STFT of the velocity-time data of each experiment are shown. The times in the figure indicate the beginning of each time window. The STFT in each experiment was implemented at a depth with a high ratio of the largest RSS of the power spectra to the RSS of the power spectra belonging to the rest of the azimuthal modes. A depth with the spherical coordinates ($r$, $\theta$)~=~(53 mm, 99\degree) fulfilled this criterion for all experiments.

Figure \ref{fig:energiesVaryingHa}(a) shows that a rotating wave with $m$~=~4 was dominant in the flow throughout the duration of the experiment. This is despite the fact that in the experiment at fixed ${\rm Ha}$ in Sec.\,\ref{sec:fixedHa}, the dominant rotating waves in the flow at ${\rm Ha} \in \{30, 32.5\}$ were $m$~=~3. On the other hand, the dominance of rotating waves with $m$~=~3 in Fig.\,\ref{fig:energiesVaryingHa}(b) and Fig.\,\ref{fig:energiesVaryingHa}(c) is not surprising, since the experiments at fixed ${\rm Ha}$ within the interval all displayed dominant RSS of the power spectra at $m$~=~3. In Fig.\,\ref{fig:energiesVaryingHa}(d), where ${\rm Ha}$ is decreased from 40 to 35, the RSS of the power spectra indicate that a rotating wave of $m$~=~2 is dominant in the flow throughout the time interval. In the fixed experiment, a rotating wave of $m$~=~3 was instead present at ${\rm Ha} = 35$. These differences in the azimuthal wavenumbers of the rotating waves at the same values of ${\rm Ha}$ depending on whether the experiments were conducted at fixed or varying ${\rm Ha}$, as seen in Fig.\,\ref{fig:energiesVaryingHa}(a) and Fig.\,\ref{fig:energiesVaryingHa}(d), demonstrate the multistable nature of rotating waves, that is, the presence of multiple solutions (distinguished by $m$) of rotating waves at certain ranges of ${\rm Ha}$. This is illustrated numerically by the solution branches of the bifurcation diagram (in both Fig.\,\ref{fig:bifurcation} and Fig.\,\ref{fig:fixedHa}), where certain ranges of ${\rm Ha}$ contain several branches. The multistability of rotating waves in our experiments is a result of differences in the initial conditions. Specifically, we see that the dynamical system tends to remain on the same solution branch it started out on even when ${\rm Ha}$ is varied linearly with time until it reaches a value of ${\rm Ha}$ where, in separate experiments conducted constantly at that value of ${\rm Ha}$, the system was attracted to a different solution branch.

\begin{figure*}
        \includegraphics[width=\textwidth]{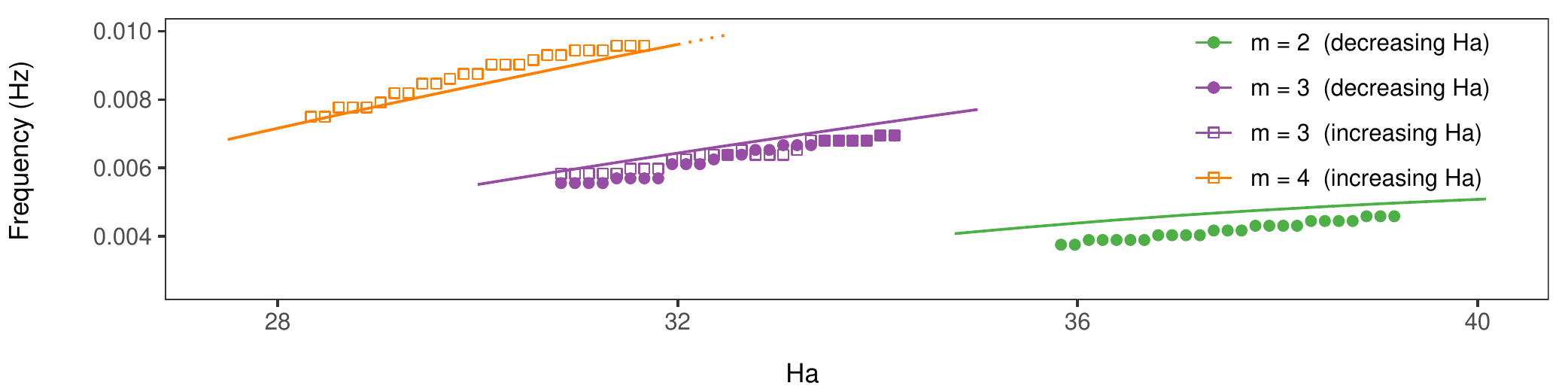}
	\caption{\label{fig:varyingHa}Rotation frequencies of the rotating waves in the flow for ${\rm Ha} \in [27.5,32.5]$, $[30,35]$, and $[35,40]$. The solid circles and the hollow squares represent the rotation frequencies obtained from the experiments with decreasing and increasing ${\rm Ha}$, respectively. The solid and dotted lines represent numerical solutions of stable and unstable rotating waves, respectively. ${{\rm Re}=1000}$ in all cases.}
\end{figure*}

Having determined the azimuthal wavenumbers of the rotating waves in the experiments with varying ${\rm Ha}$, we compute their rotation frequencies in each STFT time window. We define the temporal frequency with the largest power spectrum in each time window as the rotation frequency of the rotating wave. The results are shown in Fig.\,\ref{fig:varyingHa}. Since the ${\rm Ha}$ was linearly changed over the duration of the experiment, the value of ${\rm Ha}$ in each time window was linearly interpolated between the respective ${\rm Ha}$ intervals at the centre of the time windows. Similarly to the experiments with fixed ${\rm Ha}$ as illustrated in Fig.\,\ref{fig:fixedHa}, the rotation frequencies of the rotating waves computed from the experimental data follow the numerical solutions quite closely. In spite of the limitation in the frequency resolution, the experimental results replicate the slopes in the rotation frequencies within the respective ${\rm Ha}$ intervals. There is a significant range of overlap in the experimental rotation frequencies between $m$~=~3 and 4, once again highlighting the multistability of the rotating waves resulting from differences in initial conditions. The intention of conducting experiments at ${{\rm Ha} \in [30,35]}$, where rotating waves with $m$~=~3 was expected to be present throughout the interval, was to observe potential hysteresis in the rotation frequencies of the rotating waves depending on whether ${\rm Ha}$ was increased or decreased within the interval. However, we observed no hysteresis effect, as the rotation frequencies were almost indistinguishable from each other whether ${\rm Ha}$ was increased or decreased within the interval.

\subsection{\label{sec:initials}Influence of initially excited azimuthal modes on the multistability of rotating waves}
\begin{figure}
\includegraphics[width=\columnwidth]{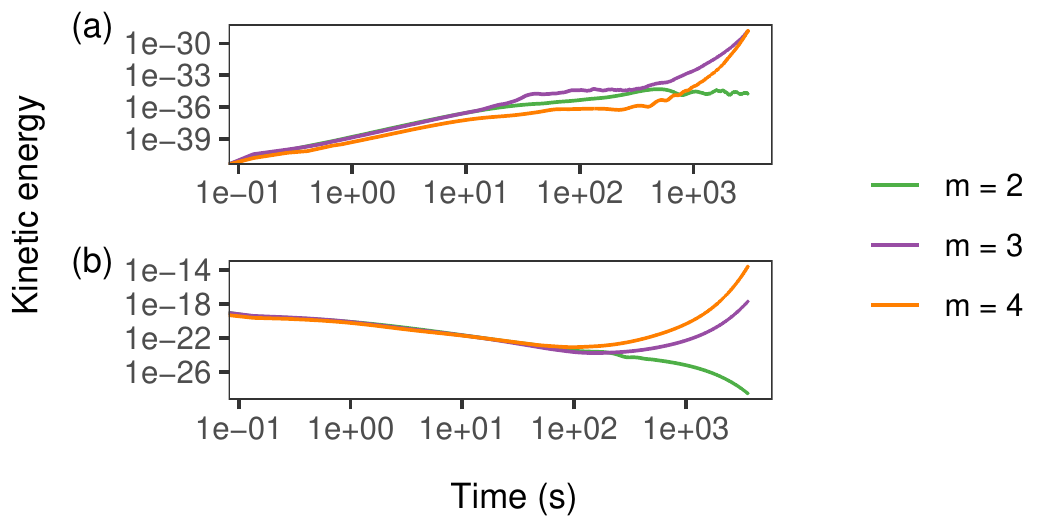}
	\caption{\label{fig:initials}Numerical simulation of the evolution of dimensionless kinetic energies possessed by rotating waves with $m \in \{2, 3, 4\}$ in the flow at ${\rm Ha}$~=~27.5 and ${\rm Re}$~=~1000 when azimuthal modes $m \in$~(a)~$\{0,1\}$ and (b)~$\{0, 1, 2, 3, 4\}$ are initially excited.}
\end{figure}

In Sec.\,\ref{sec:varyingHa}, we saw how the initial values of ${\rm Ha}$ influenced the azimuthal wavenumber of the rotating waves that formed in the MSC flow. To explore other ways in which the initial conditions affect the azimuthal symmetry of the flow, we numerically demonstrate how differences in the initially excited azimuthal modes affect the time evolution of their kinetic energies. Figure \ref{fig:initials} shows the results of this demonstration. The numerical simulation is for ${\rm Ha}$~=~27.5 and ${\rm Re}$~=~1000. Azimuthal modes $m \in \{2,3,4\}$ were analysed. In Fig.\,\ref{fig:initials}(a), only $m \in \{0, 1\}$ were initially excited, while in Fig.\,\ref{fig:initials}(b), $m \in \{0, 1, 2, 3, 4\}$ were all initially excited. Although the numerical simulations for both plots have exactly the same governing parameters, the kinetic energies possessed by the various azimuthal modes evolved quite distinctly. In Fig.\,\ref{fig:initials}(a), the kinetic energies began to gradually increase from the start. After a certain time, the kinetic energy of $m$~=~2 stabilised, while the kinetic energies in the other azimuthal modes continuously increased. This is as expected, since at ${\rm Ha}$~=~27.5, rotating waves with $m \in \{3,4\}$ are stable, while that of $m$~=~2 is unstable. The kinetic energy of $m$~=~3 was higher than that that of $m$~=~4 during the early time period. However, the kinetic energy of $m$~=~4 began to increase at a higher rate after a certain time so that towards the end of the duration, the magnitudes of the kinetic energies of $m \in \{3, 4\}$ were nearly the same. Contrary to Fig.\,\ref{fig:initials}(a), the kinetic energies of the azimuthal modes decreased at the start in \ref{fig:initials}(b). All the azimuthal modes had nearly the same magnitudes until a certain time, at which the kinetic energies of $m \in \{3, 4\}$ began to increase, with the latter increasing at a higher rate than the former. Meanwhile, the kinetic energy of $m$~=~2 continued to decrease, again demonstrating why the $m$~=~2 rotating wave is unstable at the values of ${\rm Ha}$ and ${\rm Re}$ considered.

\section{Conclusion}
We have experimentally identified rotating waves of azimuthal symmetry $m \in \{2,3,4\}$ in the return flow instability regime. The regimes where the rotating waves with the respective azimuthal wavenumbers were found corresponded reasonably well with previous numerical predictions of the stability regions of rotating waves. In addition, the rotation frequencies of the waves and the power spectra of the flow velocities obtained from the experiments showed good agreement with the numerical estimates. In the experiments where the values of ${\rm Ha}$ were varied linearly with time, the azimuthal wavenumbers of the rotating waves partly differed from those identified during experiments at fixed ${\rm Ha}$ (even at the same values of ${\rm Ha}$ and ${\rm Re}$). This experimentally demonstrates the multistable nature of rotating waves, that is, more than a single solution (distinguished by $m$) of rotating waves exist in certain regions in state space. Differences in the initial conditions and the tendency of the system to stay on the same solution branch of the bifurcation diagram play a significant role in the selection of the specific azimuthal symmetry of each saturated experimental flow.

In future, it would be worthwhile to investigate the multistability of rotating waves in the other MSC instability regimes, namely, the radial jet and the shear layer instabilities, much like we have done here for the return flow instability. Another potential area of experimental investigation would be the Hopf bifurcations of the MSC flow beyond the first bifurcation that results in the appearance of rotating waves, which may lead the flow to a state of chaos.


\begin{acknowledgments}
This project has received funding from the European Research Council (ERC) under the European Union’s Horizon 2020 research and innovation programme (grant agreement No 787544). F. Garcia was supported by the Alexander von Humboldt Foundation.
\end{acknowledgments}


\section*{Data availability}
The data that support the findings of this study are available from the corresponding author
upon reasonable request.


\renewcommand\refname{References}
\nocite{*}
\bibliography{manuscript}

\end{document}